\documentclass[fleqn,twoside]{article}
\usepackage{espcrc2}

\usepackage{graphicx}
\usepackage{epsfig}

 \newcommand{\zr}[1]{\mbox{\hspace*{#1em}}}

 \newcommand{\ZZ}{\mbox{\sf Z\zr{-0.45}Z}}

\newcommand{\AmS}{{\protect\the\textfont2
  A\kern-.1667em\lower.5ex\hbox{M}\kern-.125emS}}

\title{Electric and Magnetic Fluxes in $SU(2)$ Yang-Mills Theory}

\author{Lorenz~von~Smekal,$^a$\thanks{Talk
presented by L.~von~Smekal}  with Philippe~de~Forcrand$^b$\\
\vspace{3mm}
$^a$~Institut f\"ur Theoretische Physik III,
Universit\"at Erlangen-N\"urnberg, D-91058 Erlangen, Germany\\
$^b$~ETH-Z\"urich, CH-8093 Z\"urich and CERN Theory Division, 
CH-1211 Geneva 23, Switzerland} 

\begin{document}

\markright{\small\sf FAU-TP3-02/25, hep-lat/0209149}

\begin{abstract}
We measure the free energies in $SU(2)$  of 
static fundamental charges and center monopoles. 
Dual to temporal center fluxes, the former provide a well-defined (dis)order
parameter for deconfinement. In contrast, the monopole free energies vanish
in the thermodynamic limit at all temperatures and are thus irrelevant
for the transition.  
\vspace{1pc}
\end{abstract}

\maketitle

\section{Introduction}

For pure $SU(N)$ gauge theory without quarks 't~Hooft's gauge invariant 
electric and magnetic fluxes \cite{tHo79} describe, respectively, 
the effect of a static fundamental color charge and a center mono\-pole in a
finite volume. The partition function of a certain amount of
electric/magnetic flux yields the free energy of 
a static electric/magnetic charge with boundary conditions to imitate the
presence of its 'mirror' (anti)charge in a neighboring box along the
direction of the flux. 

Herein, we extend our previous study \cite{deF01} of the purely electric
fluxes in $SU(2)$ by measuring also the magnetic ones and combinations of the
two.

\section{Confinement: Twist vs. Electric Flux}

Imposing 't~Hooft's twisted b.c.'s fixes the
total numbers modulo $N$ of $\ZZ_N$-vortices through the 6 planes of the
4-dimensional Euclidean $1/T\! \times\! L^3$ box.~In $SU(2)$ for example, twist
in one plane corresponds to an ensemble with an odd number of
$\ZZ_2$-vortices through that plane. 
It differs by at least one from the periodic ensemble  
with an even number; and their free-energy difference is what it costs to add
one such vortex to the system. 

Intuitively it may help to assume that vortices can
lower their free energy by spreading.
At finite temperature $T>0$ it is then clear that we need to 
distinguish between the twists of two types: 

Magnetic twist is defined in a purely spatial plane in which 
the vortex can spread independent of $T$.~It fixes the conserved,
$\ZZ_N$-valued and gauge-invariant magnetic flux $\vec m $ 
through the plane. Correspondingly, its free energy, or that of a static
center monopole, will vanish for $L\!\to\!\infty$ at all $T$ as we
demonstrate in the next section. 

Twist in a temporal plane is classified by a vector $\vec k
\!\in\! \ZZ_N^3$ parallel to its spatial edge. The other edge being of finite 
length $1/T$, the vortices are squeezed in such a plane more and more 
with increasing temperature. They can no-longer spread arbitrarily 
and this is what drives the confinement phase transition. In the
thermodynamic limit, their free energy approaches zero (infinity) for $T$
below (above) $T_c$ \cite{deF01,Kov00,Sme02}. This is the reversed
behavior of a static fundamental charge.

As shown by 't~Hooft, the partition functions of fixed units of electric and 
magnetic fluxes, $\vec e, \vec m \!\in\! \ZZ_N^3$, which we denote 
by $Z_e(\vec e, \vec m)$, are obtained as 3-dimensional $\ZZ_N$-Fourier
transforms,  w.r.t. the temporal $\vec k$-twist, of those with twisted b.c.'s,
$Z_k(\vec k,\vec m)$.~With the free energy of a purely electric flux one
measures that of static fundamental charge in a perfectly well-defined
(UV-regular) way \cite{deF01}.~This follows from the gauge-invariant
definition of the Polyakov loop  $P(\vec x)$  in presence of temporal twist
which entails, 
\begin{equation}
 Z_e(\vec e,0) = 
        \big\langle P(\vec x) P^\dagger(\vec x +L\vec e) \big\rangle _{L,T}
        \; , 
\label{Eq:1}
\end{equation}
relative to the no-flux ensemble ({\it i.e.}, the expectation value is taken
therein and the l.h.s.~is normalized such that $Z_e(0,0)\!=\!1$).
In Ref.~\cite{deF01} we measured the ratios of partition functions 
$Z_k(\vec k,0)$ with different $\vec k$-twists
and the $Z_e(\vec e,0)$ for the various electric fluxes in $SU(2)$.
We explicitly demonstrated the Kramers-Wannier duality, 
between the temporal center fluxes and the static charges, 
of the general pattern 
\begin{equation}
Z_e(\vec e,0) \,
        \stackrel{L\to\infty}{\longrightarrow} \,  \left\{ { 
                 \, 0 \, , \;\; Z_k(\vec k,0) \!\to\! 1 , \; T < T_c
                                                   \atop 
                  \, 1 \, , \;\; Z_k(\vec k,0) \!\to\! 0 , \; T > T_c
                            } \right.  
\end{equation}
reflecting the different realization of the 3-dim. (electric) center
symmetry in both phases. As expected from universality, this duality 
is the analogue in $SU(2)$ of that between the Wilson loops of the 3d 
$\ZZ_2$-gauge theory and the 3d-Ising spins.

\begin{figure}[t]
\vspace{-8pt}
\epsfig{file=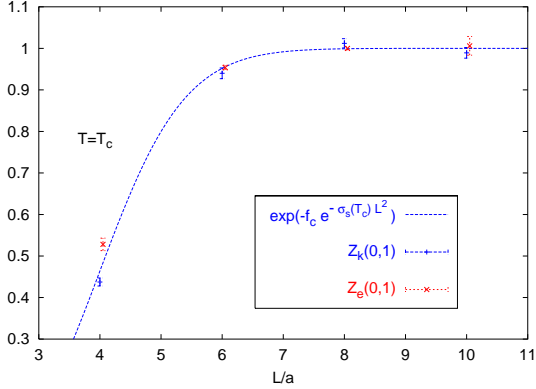,width=\linewidth}
\vspace{-1.5cm}
\caption{Finite volume partition functions of one magnetic flux at $T_c$,
$Z_k(0,1)$ and $Z_e(0,1)$ relative to the periodic and no-flux ensemble,
respectively.}   
\label{Zbc}
\vspace{-.5cm}
\end{figure}

\section{Magnetic Flux}

There is no analogue in the 3d-spin systems to the magnetic fluxes, however.
These are temporal 't~Hooft loops of maximal size winding on the dual lattice
around the temporal $1/T\!\times\! L$ planes. 
The magnetic current along the 't~Hooft loop in this case creates a pair of
static center monopoles at a distance $L$. 
Magnetic flux $\vec m$ thus corresponds to one static center monopole
inside the $L^3$ box with mirror image in a neighboring box along the
direction of $\vec m$ -- the magnetic analogue of the Polyakov loop
correlator on the r.h.s.~of Eq.~(\ref{Eq:1}).

\begin{figure}[t]
\vspace{-8pt}
\epsfig{file=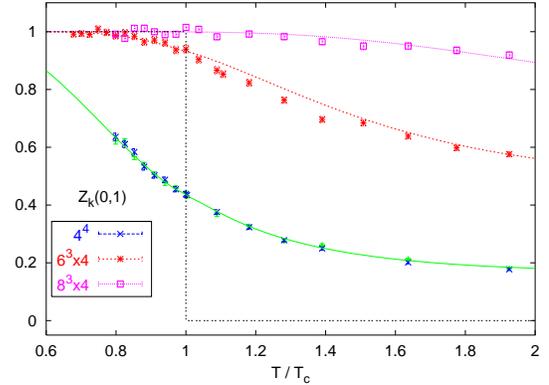,width=\linewidth}
\vspace{-1.5cm}
\caption{Magnetic flux versus temperature.}
\label{Zkm}
\vspace{-.5cm}
\end{figure}

In agreement with this electric-magnetic duality, 
the temporal 't~Hooft loops in $SU(2)$ show a screening behavior 
in both phases, for $T\!<\!T_c$ and for $T\!>\!T_c$ \cite{Kov00,deF00}, just
as spatial Wilson loops exhibit an area law in either case. 
One can thus anticipate likewise that the monopole free
energy in both phases must vanish in the thermodynamic limit. 
This would not exclude that it approaches a finite value at $T\!=\!T_c$ as in
the case of electric charges and temporal center fluxes. We obtained in
Ref.~\cite{deF01}, {\it e.g.}, with $\vec k\! =\! (0,0,1)$ for one unit of the
latter, $Z_k(\vec k,0)= 0.54(1) $ at $T_c$ in agreement with the
universally related ratio in the Ising model.~In particular, if the massless
phase which the system passes through at $T_c$ was 
selfdual, one would also expect a finite monopole free energy at $T\!=\!T_c$.
This is, however, not the case, as shown in
Fig.~\ref{Zbc}. The free energies of the magnetic fluxes in $SU(2)$ at
$T_c$ are well described by an exponential decrease $\propto 
\exp\{-\sigma_s(T_c) L^2\} $, where $\sigma_s $ is the spatial string tension. 
As a check, our fit then yields for the spatial string tension at
$T_c$, 
\[
  \sigma_s = (2.2 \pm 0.2) \, T_c^2\,, \;\mbox{or} 
   \;   T_c/\sqrt{\sigma_s} = 0.675 \pm 0.03 \; , 
\]
consistent with published values for the zero temperature  $SU(2)$
string tension \cite{Fin92}.~This screening of the temporal
't~Hooft loops is similar to that of spatial ones
in the low temperature phase: 

In \cite{deF01} we demonstrated that
$-\ln Z_k(\vec k,0) \propto \exp\{-\sigma(T) L/T\}$ 
for temporal $\vec k$-twist at $T\!<\!T_c$ and sufficiently large $L$.~The
important difference is the temperature dependence of the standard
string tension $\sigma(T)$, as compared to the practically constant  
spatial one, $\sigma_s$ here.~As a result, 
the dominant behavior of the monopole free energies cannot be described by
finite size scaling. 

That they 
indeed vanish for all other $T$ also, is seen in
Fig.~\ref{Zkm} where we plot the partition function $Z_k(0,\vec m)$ 
of one magnetic twist ({\it e.g.}, $\vec m \!=\! (0,0,1)$)
versus temperature. While it is indistinguishable from a temporal
twist on the symmetric $4^4$ lattice as expected, 
with increasing spatial lattice size it rapidly and smoothly approaches
unity over the whole temperature range of our simulations.

\begin{figure}[t]
\vspace{-8pt}
\epsfig{file=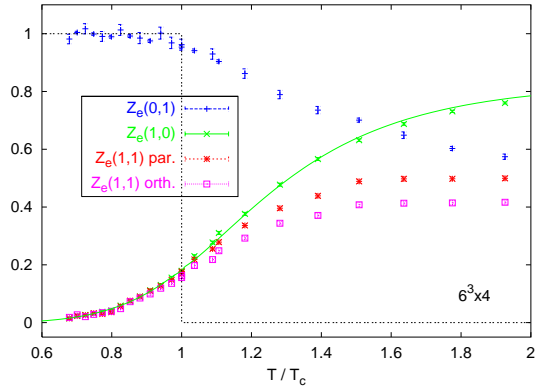,width=\linewidth}
\vspace{-1.5cm}
\caption{Electric and magnetic fluxes  
on the $6^3\!\times \!4$ lattice versus temperature.}    
\vspace{-.5cm}
\label{Zem6}
\end{figure}

\begin{figure}[t]
\vspace{-8pt}
\epsfig{file=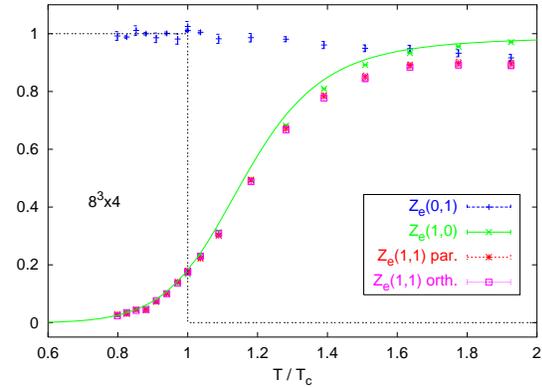,width=\linewidth}
\vspace{-1.5cm}
\caption{Same as in Fig.~\protect\ref{Zem6} for the $8^3\! \times \!4$
lattice.}    
\vspace{-.5cm}
\label{Zem8}
\end{figure}

Our measurements of the flux partition functions $Z_e(\vec e,\vec m)$
from two different lattice sizes are shown for various combinations of
magnetic with electric fluxes in Figs.~\ref{Zem6} and \ref{Zem8}.  
The differences between parallel ($\vec e\cdot\vec m$ odd) and orthogonal
($\vec e\cdot\vec m$ even) fluxes on the smaller volumes have previously been
observed at $T\!=\!0$ also~\cite{Gon96}. 
They were then found in good agreement with 
semiclassical predictions. As the purely magnetic free energy 
vanishes, or $Z_e(0,\vec m)\!\to\! 1$ with
increasing spatial size, these differences disappear and the
partition functions approach those of purely electric fluxes for which we
include the fit from \cite{deF01} in the figures. 
Similarly, our data for all other combinations of twists and fluxes in
$SU(2)$ indicate, at all  $T$,
\vspace{-.1cm}
\[ Z_e(\vec e,\vec m)  \stackrel{L\to\infty}{\longrightarrow}  Z_e(\vec e,0)
\, ,    \;\;    Z_k(\vec k,\vec m) 
              \stackrel{L\to\infty}{\longrightarrow} Z_k(\vec k,0).
\]
\vspace{-.1cm}
Magnetic fluxes are irrelevant for the phase transition, and center
monopoles always 'condense'. 
The corresponding 3-dim. magnetic center symmetry remains unbroken at all
temperatures.

\section{Conclusions}

To summarize, changing the spatial twist is a gauge-invariant way of
introducing one more static center monopole.
The monopole free energy in the 3d Georgi-Glashow model 
was recently studied in an analogous way \cite{Dav02}. 
As therein, we observe for $SU(2)$ that the monopole free energy is zero, in
the thermodynamic limit, at all temperatures. In contrast to the crossover 
behavior of the Georgi-Glashow model, however, in $SU(2)$ there is no
indication of a plateau with finite monopole mass at intermediate volumes
either. 

Simulations were performed on the SGI Origin systems at RRZE, Erlangen, and
ZHR, Dresden.  

\markright{}

\end{document}